\documentclass{article}
\usepackage{amsmath,amsthm,amssymb}
\usepackage{color,soul}
\usepackage{float} % advanced positioning of figures and tables
\usepackage{graphicx} % including graphics
\usepackage[hidelinks]{hyperref}
\usepackage{bookmark} % creates PDF bookmarks
\usepackage{subfigure} % allows for two figures to be included in one figure
\usepackage{natbib} % customizes citations in the text
\setcitestyle{aysep={}} % removes commas in references
\usepackage{booktabs} % Allows use of \toprule for tables
\usepackage{chngpage} %temporary adjusting of margins

\usepackage{ragged2e} %allows for justify to work in tables

\makeatletter
\renewcommand\hyper@natlinkbreak[2]{#1} % combines in-text reference links from 2 to 1
\makeatother
\usepackage{bbm}

\usepackage{epstopdf} % ensures that eps files are converted to PDF so Windows doesn't freak,
\usepackage{mdwlist}
\usepackage{pdflscape}
\usepackage[section]{placeins} % keep figures within section
\usepackage{caption} %add captions below figures (use with below)
\usepackage{pgfplots}
\usepackage{forest} % Helps produce Quadrilha figure structure
\pgfplotsset{compat=1.18}

\usetikzlibrary{backgrounds, arrows, shapes, shapes.geometric, shapes.misc}
\tikzstyle{tikzfig}=[baseline=-0.25em,scale=0.5]
\tikzstyle{none}=[inner sep=0mm]
\usetikzlibrary{decorations.pathreplacing} % for Game Tree
\tikzset{
solid node/.style={circle,draw,inner sep=1.5,fill=black},
hollow node/.style={circle,draw,inner sep=1.5}
}

\title{AI and the law}

    \author{Henry A. Thompson\footnote{I would like to thank Francesco Parisi for encouragement. Department of Economics, University of Mississippi, hathomp@olemiss.edu.}}

\date{}

\begin{document}
\maketitle
\begin{abstract}
    I argue that generative AI will have an uneven effect on the evolution of the law. To do so, I consider generative AI as a labor-augmenting technology that reduces the cost of both writing more complete contracts and litigating in court. The contracting effect reduces the demand for court services by making contracts more complete. The litigation effect, by contrast, increases the demand for court services by a) making contracts less complete and b) reducing litigants' incentive to settle, all else equal. Where contracts are common, as in property and contract law, the change in the quantity of litigation is uncertain due to offsetting contracting and litigation effects. However, in areas where contracts are rare, as in tort law, the amount of litigation is likely to rise. Following \cite{rubin1977common} and \cite{priest1977common}, generative AI will accelerate the evolution of tort law toward efficiency.
\end{abstract}
\newpage

\section{Introduction}

Scholars have long recognized that technological progress shapes the evolution of the common law. For example, the proliferation of rail and heavy machinery during the Industrial Revolution spurred the shift from strict liability to negligence by increasing the number and severity of accidents \citep{posner1972theory, gifford2018technological}.\footnote{\cite{fleck2024courts} also study the railway boom. They find that the growth of rail encouraged a shift away from a property rule called the benefit offset that helped railroads reduce payments for land acquired through eminent domain.} In the 20th century, increases in the complexity of production drove the rise of strict products liability \citep{landes1989economic}. Technological improvements at home and at work propelled the no-fault divorce revolution of the 1970s by increasing women's returns on the labor market \citep{leeson2017economic}.\footnote{See also \cite{friedman1984rights}.} Such work shows that whenever new tech switched the least-cost avoider or highest-valued user of a property right, the law evolved with it. 

I consider how the law will change in response to an altogether different kind of innovation. Generative AI is a labor-augmenting technology that is especially likely to change the costs of litigation rather than the least-cost avoider. Indeed the most striking aspect of generative artificial intelligence (AI) is how much it boosts user productivity. Preliminary research has found that generative AI boosts materials discovery by 44\% \citep{toner2024artificial}, programming speed by 56\% \citep{peng2023impactaideveloperproductivity}, customer support agent speed by 14\% \citep{brynjolfsson2023generative}, writing task completion speed by 40\% \citep{noy2023experimental}, and consulting output and quality by 12\% and 40\% \citep{dell2023navigating}. \cite{choi2023lawyering} find that generative AI boosts legal task completion speed by 12-32\%. In this article I explore the implications of such advances for the law.

I argue that generative AI will alter the speed of the law's evolution in two ways. First, generative AI will help attorneys write more complete contracts, weakening the demand for courts. Second, generative AI will reduce the cost of litigation, thereby boosting the demand for courts. When inefficient rules are litigated more often than efficient ones, these changes in the demand for courts will change how quickly the law evolves \citep{priest1977common, rubin1977common}. 

At first glance, the net effect of generative AI on the law’s rate of change \textit{in totum} is ambiguous. The two effects offset one another. However, each effect will not be present in every area of law. 

In tort law, legal change will accelerate. This is because there are no contracts that would reduce the demand for courts. As a result, cheaper litigation will increase the rate at which inefficient allocations of tort liability are challenged and ultimately overturned \citep{priest1977common, rubin1977common}.

In property and contract law, the contracting effect and the litigation effect work in opposite directions. Although generative AI incentivizes more litigation due to lower litigation costs, it also incentivizes less litigation thanks to more complete contracts. The net effect depends upon how elastic the marginal benefit of gap-filling is relative to the marginal cost of gap-filling.

The demand for courts and legal change is likely to fall on net where the marginal benefit of contractual gap-filling is more elastic. Examples include commercial real estate and major infrastructure projects where even minor problems can have uniformly high costs. However, the demand for courts and legal change is likely to rise on net where the marginal costs of contractual gap-filling are more elastic as in commodity trading markets, mass consumer goods, and markets in which most participants are retired.

My article contributes to three branches of scholarship. The first branch studies technology-driven institutional change. There, scholars have studied how technology sparks changes in the organization, structure, and scope of government \citep{batchelder1983rational, allen1998compatible, allen2011institutional, allen2011evolution} or religious institutions \citep{rubin2014printing}.\footnote{Other papers have studied how institutions slow or speed up the adoption of novel technologies \citep{cocsgel2012politicaleconomy, allen2015institutionally}.} The second is the now substantial body of scholarship on the evolutionary efficiency of law inculcated by \cite{priest1977common} and \cite{rubin1977common}.\footnote{See, for instance, \cite{parisi2001genesis, parisi2002entropy, fon2003litigation, fon2006judicial, albrecht2022evolution}.} The paper also contributes to the growing body of work on AI in the law. The predictive ability of AI features prominently there \citep{alarie2018artificial, gans2024demand}.\footnote{For work that examines the regulation of AI, see for instance \cite{arbel2024systemic}} Of the work on AI in the law, the papers nearest to my own are \cite{gans2024demand} and \cite{casey2020will}. \cite{gans2024demand} examines the demand for AI when it can produce accurate predictions about the outcome of a trial. \cite{casey2020will} evaluate the effect of ``robot judges" on litigation and settlement outcomes. Neither consider the implications of generative AI for contract completeness, attorney productivity, or the evolution of the law.

\section{Context}

Attorneys produce and sell information about property rights.\footnote{For a slightly different perspective, see \cite{gilson1984value} who suggests that a business attorney's main purpose is to reduce information asymmetries between the two parties.} For example, before a dispute occurs and where a contract exists, attorneys produce information about the seller's current assets and liabilities and a buyer's rights and future obligations. Attorneys also produce and sell information after a dispute has occurred, whether a contract exists or not. In that case, they sell information about the law, regulations, precedents, sound legal arguments, and procedural rules that are beyond their clients' ken.

To produce information about property rights, an attorney spends a great deal of time either reading or writing. The three main tasks of an attorney are legal search, creating legal arguments, and legal writing \citep{tu2023artificial}. Legal search involves an attorney learning about the law relative to the facts of the case at hand. An attorney then creates a legal argument that either highlights the case's similarities or differences relative to precedent. Legal writing then involves creating a memo or brief that ``highlights the key facts, issues, legal rules, and public policy arguments is crucial for producing effective descriptive and advocacy work" \citep[p. 106]{tu2023artificial}. Each task involves a great deal of reading, writing, or both. 

Given how much time attorneys spend reading and writing, generative AI is especially likely to change the practice of law. In November 2022 the lab OpenAI released a large language model (LLM) called ChatGPT. The tool was quite adept at producing large quantities of useful and meaningful text, the attorney's \textit{forte}. Two months after its release, ChatGPT had 100 million users. 

LLMs have grown quickly in number and advanced rapidly in quality since then. By the summer of 2023, the popular open-source platform Open Arena had around 25 models that users could try out \citep{zheng2023lmsys}. As of November 2024, there were 161 different LLMs including Google's Gemini, xAI's Grok, Anthropic's Claude, Meta's Llama, and Microsoft's Copilot \citep{chiang2024chatbot}, a 644\% increase in the number of models in about a year.

Such generative AI tools have been adopted quickly. Bick et al. found that among the U.S. adult population, ``Generative AI has a 39.5 percent adoption rate after two years, compared with 20 percent for the internet after two years and 20 percent for PCs after three years" \citeyearpar[2]{bick2024rapid}. Bick et al. did find that legal occupations had relatively lower rates of adoption as of August 2024. That, however, is likely to change \citep{tu2023artificial, arbel2024generative, arbel2024judicial}.

Tu et al. suggest that generative AI will help attorneys with legal search, arguments, and writing by identifying the key similarities and differences with the black letter of the law, generating new legal arguments, and drafting memos and briefs. \citet[31]{villasenor2023generative} too expects Generative AI to ``accelerate legal research and to produce drafts of text for use in contracts, regulatory filings, court rulings, academic papers, wills, trusts, patent specifications, affidavits, articles of incorporation, and more."\footnote{\citet{mcginnis2013great} argue that machine intelligence is likely to influence five services provided by attorneys: discovery, legal search, document generation, brief generation, and the prediction of case outcomes.}

As a labor-augmenting technology, generative AI will have two important, economic effects on attorney work. First, generative AI is likely to make attorneys more productive as contract drafters by reducing the cost of a) writing a contract \textit{ex nihil} and b) making any existing contract more complete. Early research already suggests that generative AI helps users search, summarize, outline, and draft documents of many kinds, including contracts. I dub this the ``contracting effect."
Second, generative AI is likely to make trial attorneys more productive. This is because, as discussed above, generative AI helps attorneys to do trial work far more quickly (preparing legal arguments, filing motions and briefs, and preparing opening and closing statements far more quickly), reducing the marginal cost of providing such services. The result is a lower equilibrium price of buying the services of a trial attorney. I refer to the reduction in the price of trial attorney services as the ``litigation effect."

Generative AI's two economic effects will become legal ones via the courts. If the contracting effect and litigation effect change the demand for courts, then the law's evolution will reflect that.

There is good reason to think that the contracting effect and litigation effect will change the demand for courts. Courts are some of the most important suppliers of formal dispute resolution available \citep[218]{farnsworth1990}. People use them as ``\textit{ex post}" gap-fillers and resolvers of tortious disputes.\footnote{U.S. courts supply other services too. Among other things, they resolve disputes, interpret and enforce explicit contractual terms, and help parties rewrite terms within a contract.} As gap-fillers, courts aim to resolve a contractual dispute in a way that does not violate the contract's explicit terms \citep[343, 349-350]{farnsworth1990}. Such a service is backward-looking, aiming to infer what the litigants ``would have wanted" had they been able to anticipate the realized state of the world. For tortious events, courts aim to resolve disputes by determining harm, cause, and fault.

The demand for courts as an \textit{ex post} gap-filler depends on, for example, how complete contracts are and the costs of litigation. Generative AI may impact the demand for courts by changing parties' incentives before a dispute occurs and then once a dispute has occurred. To evaluate these effects, I employ a few canonical models in Law \& Economics.

\section{AI and the contracting decision}
Generative AI is likely to change how people contract with one another. On the one hand, the contracting effect makes contracts more complete by better anticipating different states of the world and better delineating property rights therein. On the other hand, the litigation effect will make contracts less complete by reducing the cost of resolving disputes \textit{ex post}. When lawyers are cheaper to hire, parties might opt to economize on \textit{ex ante} contracting costs and instead resolve disputes \textit{ex post}.

To see which effect dominates, meet Quincy who wants to buy a van from Genevieve. He can buy a new van or a used van. The new van is a ``safer" purchase because it is unused. As a result, the new van commands a high price. The opposite is true of the used van, whose lower price reflects the greater risk of purchase.

In a world without generative AI, Quincy might just buy the new van. In a world with generative AI, however, Quincy might instead opt to buy the used car despite its greater risk. This is true because generative AI may increase the completeness of contracts, reducing the risk of buying a used van.

\subsection{A model of incomplete contracts}
\begin{figure}
    \centering
    \includegraphics[width=0.8\linewidth]{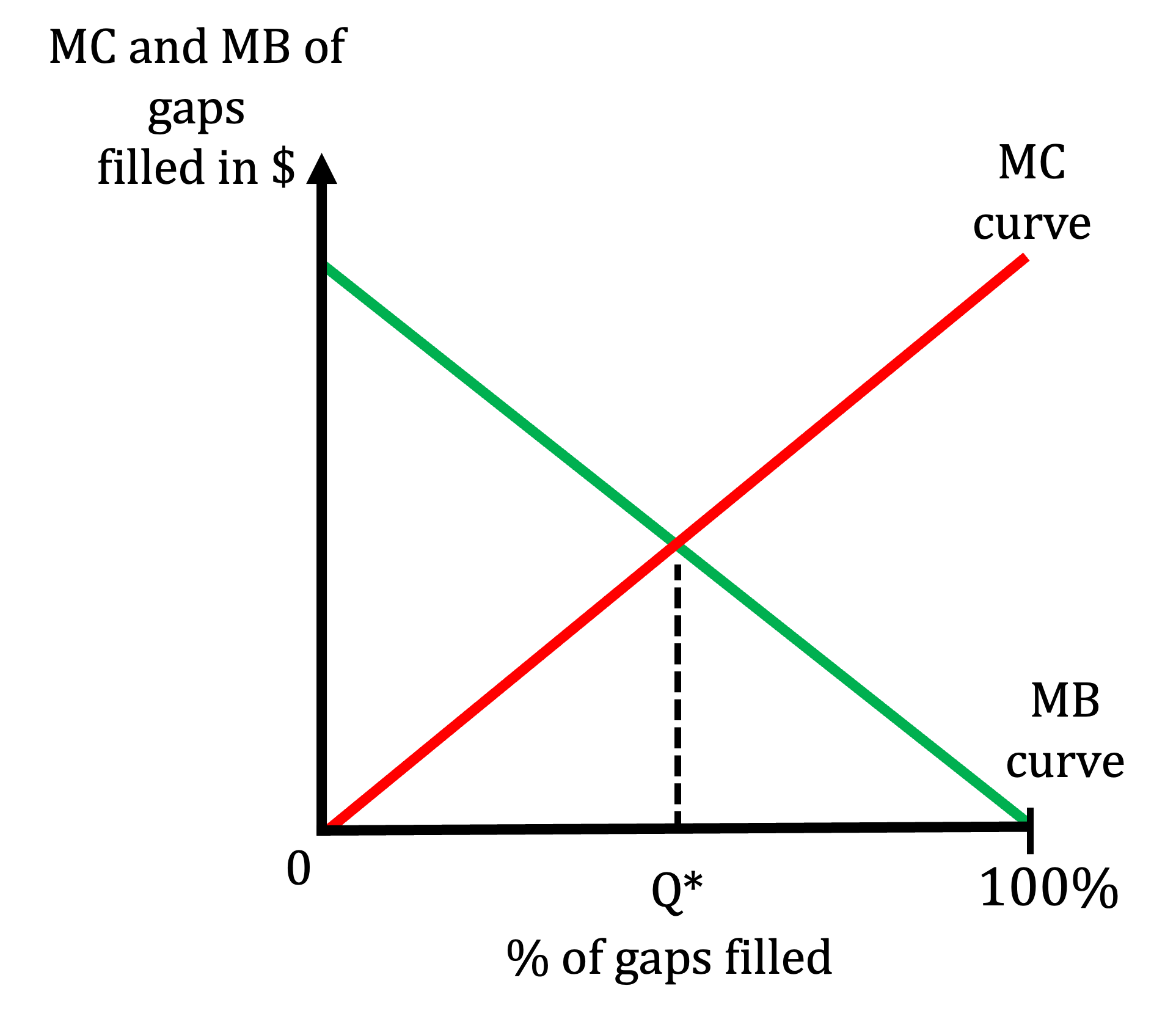}
    \caption{Equilibrium contract completeness}
    \label{fig:inccontract}
\end{figure}

Contracts delineate property rights in different future states of the world. Contractual contingencies give, for example, Quincy the right to a new battery if it dies within 30 days or Genevieve the right to seize the car if Quincy falls behind on his payments. Contracts are useful to the pair because contracts reduce the costs of \textit{ex post} dispute resolution.

When buying a used van, Quincy has two options. On the one hand, he can delineate property rights upfront by hiring an attorney to write a relatively complete contract that will account for the many possible problems with the used car. Dealing with the risks upfront is beneficial because it means Quincy (and Genevieve) saves money that he might otherwise have to spend resolving disputes later if they occur. Making a contract more complete \textit{ex ante} reduces how much must be spent resolving disputes \textit{ex post}.

On the other hand, Quincy can plan to delineate rights once problems arise. In this case, Quincy avoids the cost of hiring an attorney upfront. The cost is that Quincy is more likely to have to hire a trial attorney later once a dispute occurs.

\textit{Ceteris paribus}, Quincy benefits from making his van contract more complete. However, the added benefit he gets from filling a gap in the contract is decreasing as in Figure \ref{fig:inccontract}. These marginal benefits decrease because Quincy and Genevieve address the most important and likely issues first, leaving less critical gaps for later. As more gaps are filled, the remaining gaps become less significant. Thus the marginal benefit of filling an extra percent of gaps is decreasing as more gaps are filled.

Making a contract more complete is costly. Quincy's attorney must spend time, legal fees, and effort to foresee and specify any possible scenario. Indeed, each additional contingency is more costly to add than the previous one because initial gaps are easier to identify and address. Successive contingencies become more and more resource-intensive as they become increasingly complex and time-consuming to fill. The rising opportunity cost of gap-filling implies that Quincy's marginal cost of filling an extra percent of gaps is rising as more gaps are filled, as shown in Figure \ref{fig:inccontract}.\footnote{There are at least two other costs that can prevent most contracts from being complete. First, parties to a transaction have an incentive to spend resources verifying \textit{ex post} what has happened. Second, enforcement costs matter too. Even if parties to a transaction can cheaply learn what will or has happened, they must still be able to enforce their property right. Given the ability of generative AI to produce useful and meaningful text, generative AI is unlikely to impact either one.}

The decreasing benefits and rising marginal costs of gap-filling mean that Quincy faces a tradeoff when deciding how complete to make his contract. Quincy must balance the benefits of fewer \textit{ex post} disputes against the costs of predicting and specifying each possible future event. As a result, Quincy will increase the relative completeness of his contract until the marginal benefit of the gaps filled is just equal to the marginal cost thereof. As reflected in Figure \ref{fig:inccontract}, Quincy's equilibrium contract is likely to be incomplete. 

\subsection{Comparative statics}
\begin{figure}
    \centering
    \includegraphics[width=1\linewidth]{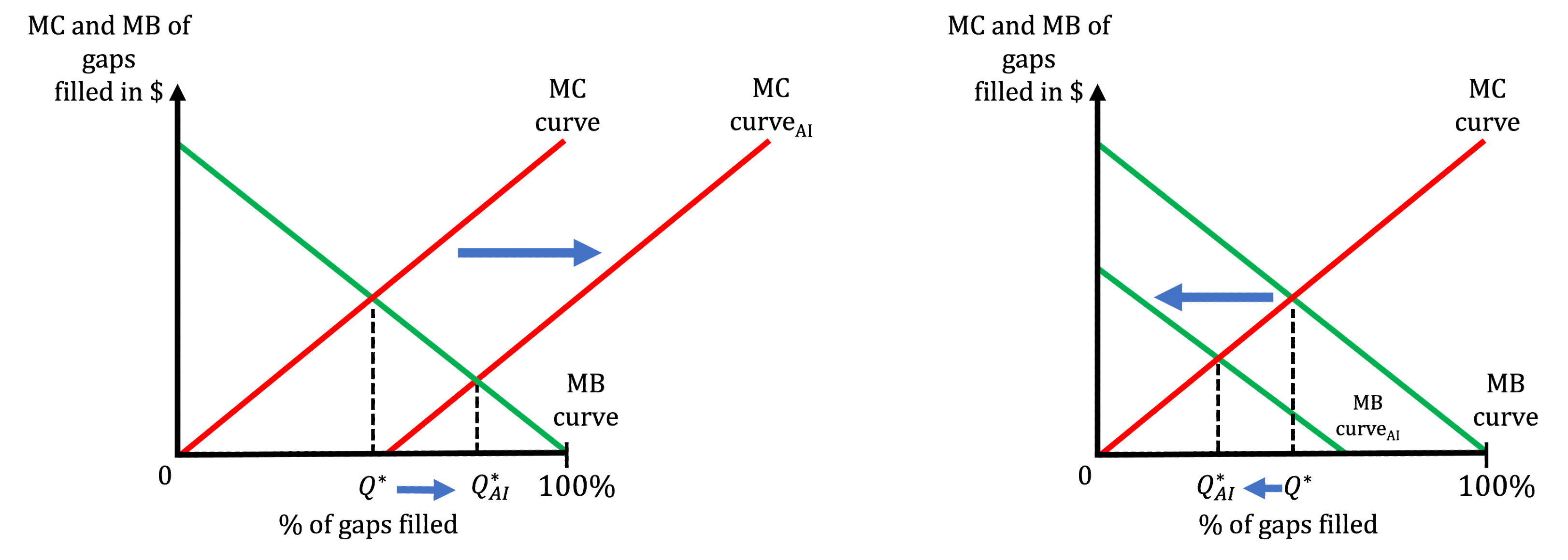}
    \caption{Some comparative statics}
    \label{fig:contractingcomparative}
\end{figure}

Generative AI will have two offsetting effects on just how incomplete Quincy's contract is. Generative AI will make his contract more complete by reducing the cost of gap-filling. This is because it makes attorneys more productive and effective contract drafters by automating drafting and review, drafting longer and more complex clauses, and identifying gaps that might otherwise go unseen. As shown in the first panel of Figure \ref{fig:contractingcomparative}, generative AI will cause the marginal cost of gap-filling to fall, shifting the marginal cost curve to the right and leading to a higher fraction of gaps filled in equilibrium. 

Generative AI will also reduce the benefits of making contracts more complete. This is true because hiring an attorney is one of the chief costs of \textit{ex post} dispute resolution. Generative AI reduces that cost due to the ``litigation effect." Because generative AI also boosts an attorney's productivity with respect to litigation, their services are cheaper to hire and the cost-savings that Quincy can expect to earn from making a contract more complete are lower. So by reducing the cost of \textit{ex post} dispute resolution, generative AI reduces the marginal benefits of making a contract more complete, implying a leftward shift in the marginal benefit curve and a less complete contract in equilibrium. This effect is shown in the second panel of Figure \ref{fig:contractingcomparative}.

The contracting effect and litigation effect offset one another. The net effect of generative AI on the equilibrium amount of contract completeness is, therefore, ambiguous. Which effect dominates will depend upon a) the relative elasticities of the marginal cost and marginal benefit curves, and b) the relative magnitudes of the contract effect and the litigation effect. I examine each briefly.

Consider first the relative elasticities of the marginal cost and benefit curves of contractual gap-filling. A leftward shift in the marginal benefit curve due to the litigation effect may decrease the degree of contract completeness a lot or a little. A more elastic marginal cost of gap-filling will cause the decline in the degree of contract completeness to be large. By contrast, a less elastic marginal cost curve will cause the decline to be small.

There are at least two important sources of greater elasticity for the marginal cost curve: input variability and issue complexity.\footnote{If issues are largely self-contained, the marginal cost curve will be more elastic as adding new terms needs little restructuring of other parts of the contract when additional terms are added.} The marginal cost of filling contractual gaps will be more elastic when the resources such as time and effort used to make the contract more complete are more homogeneous. In that case, the resources used to fill initial gaps are just as suitable as resources used to fill later gaps, implying a more elastic marginal cost curve. Inputs for gap-filling are more likely to be homogeneous in areas where, for example, workers have lower wages, lower demands on their time, and are more knowledgeable about the law. %

The marginal cost curve will also be more elastic when the good to be traded is simple and varies little in its valuable properties. For example, the marginal cost curve for gap-filling is more elastic for mass-produced goods than for custom-made ones. When goods vary little in valuable attributes, there is little variance for a contract to account for. The first contingencies are just as easy to include as the last ones.

A reduction in the marginal cost of gap-filling will make contracts more complete, but the size of the effect may be large or small. The size of the change depends upon the relative elasticity of the marginal benefit curve. If the marginal benefit of the gap-filling curve is elastic, the increase in contract completeness will be large. If the marginal benefit curve is inelastic, the change will be small.

Consider two important sources of greater elasticity for the marginal benefit curve. First, if problems solved via gap-filling are uniform, then the marginal benefits fall slowly with greater completeness. This is because the cost-savings of gaps that are filled first are similar to the cost-savings of gaps filled later. In a high-stakes construction project, for example, any single delay or error will entail high dispute costs, implying a more elastic marginal benefit curve. When stakes are highly variable, marginal benefits will decrease more steeply, as in insurance. 

Second, for repeated trades, the marginal benefit of filling contractual gaps falls quickly. This is because parties have other, informal ways to resolve disputes, including renegotiation. By contrast, for one-time trades, there is no such option. As a result, the marginal benefit of adding additional stipulations to a contract falls less quickly. We can expect long-term relationships between wholesalers and retailers to meet this condition compared to, for example, spot markets. 

The size of the attorney and contracting effects may also vary. The litigation effect may be large or small. The larger the size of the attorney effect, the more the marginal benefit curve will shift to the left, and the less complete Quincy's contract will be. The contracting effect may also be large or small. The larger the size of the effect, the further to the right the marginal cost curve will shift and the more complete Quincy's contract will be.

It is not clear how large these effects will be. For the sake of simplicity, I assume that the effects are comparable in size. As a result, the net effect will depend only upon the relative elasticities of the marginal cost and benefit curves.

\subsection{Implications for courts}
Contracts and courts are substitutes. Contracts help to delineate rights \textit{ex ante}, whereas courts help to define rights \textit{ex post}. As a result, the demand for courts depends upon how complete contracts are. If generative AI causes contracts to become more complete, the demand for courts will fall. But if generative AI causes contracts to become less complete, the demand for court-provided dispute resolution will rise.

If the contracting effect dominates, then contracts will become more complete and litigation will become less common. This may occur in commercial real estate development and major infrastructure developments. In both, the complexity of the stipulations makes the marginal cost curve less elastic and the prospect of uniformly high costs of delays makes the marginal benefit curve more elastic. Ergo, participants will substitute away from formal methods of dispute resolution as the cost of making contracts more complete falls. 

If the litigation effect dominates, then contracts will become less complete and litigation will be more common. This may occur in commodity trading markets, mass consumer goods, and markets in which most of the participants are retired. In the first two, goods are more homogenous. In markets where most participants are retirees, the opportunity cost of adding more contingencies rises slowly. In all three areas, marginal cost curves are likely to be more elastic, implying that participants substitute away from more complete contracts into formal methods of dispute resolution.

\section{AI and the settlement decision}
\label{sec:settlementdecision}

In addition to the contracting decision, generative AI is likely to impact the settlement vs. trial decision. 

\subsection{A model of settlement}
Suppose that Quincy's van (which he bought and affectionally names the ``swagger wagon") explodes. Quincy suspects that the car dealer, Genevieve, was careless and so wants to recover his costs whether through settlement or at trial. Genevieve, by contrast, wants to minimize her costs.\footnote{The model below is canonical in Law \& Economics \citep{landes1971economic, gould1973economics, shavell1982suit}.}

Quincy can try to settle or go to trial. If Quincy offers to settle with Genevieve and Genevieve agrees, Quincy gets a benefit the value of which is $S$. Genevieve by contrast pays $S$. I assume that the costs of settling are zero.\footnote{Settlement agreements are already simple documents, so AI will only slightly improve their completeness. This small improvement matters even less when compared to the much larger costs of negotiating the settlement amount. And both the drafting and negotiation costs of settling are far lower than the costs of going to trial \citep{cooter1989economic}. Given these relative costs, the contracting effect will have minimal impact on the decision to settle relative to the litigation effect.} 

If, however, Quincy goes to trial and wins he gets the judgment of size $J$ less the cost of going to trial, $C_Q$ or $J-C_Q$. The cost of going to trial includes, for example, attorney and court filing fees. Quincy's gain is Genevieve's loss, so when Quincy wins Genevieve must pay both the judgment $J$ and her costs of going to trial: $J+C_G$.

If Quincy goes to trial and loses, Quincy pays the cost of going to trial, $C_Q$, and gets no benefit. In that case, Genevieve only has to pay the cost of going to trial, $C_G$.

Quincy thinks that his chance of winning if he goes to trial is $P_Q$. Quincy's expected benefit of going to trial is therefore
\begin{align*}
    P_Q(J-C_Q)+(1-P_Q)(-C_Q)
\end{align*}
or
\begin{align}
   P_QJ - C_Q.
\end{align}

Conversely, Genevieve thinks that the chance that she loses liability at trial is $P_G$, and the probability that she escapes liability is $1-P_G$. Thus her expected cost of going to trial is
\begin{align*}
    P_G(J+C_G)+(1-P_G)(C_G)
\end{align*}
or
\begin{align}
    P_GJ + C_G.
\end{align}
Quincy will want to settle when the benefit of settlement is greater than or equal to the expected benefit of going to trial: $S \geq P_QJ - C_Q$. Quincy will, by contrast, go to trial when the expected benefit of going to trial exceeds the benefit of settling: $P_QJ - C_Q > S$. 

By contrast, Genevieve will try to settle when the cost of settlement is lower than the expected cost of going to trial: $P_G J + C_G \geq S$. Genevieve will opt for a trial instead when the expected cost of the trial is lower than the cost of the settlement: $S > P_GJ + C_G$.

Ergo, settlement will occur as long as $S$ is higher than Quincy's expected benefit of trial and is lower than Genevieve's expected cost of going to trial:
\begin{align}
    P_QJ - C_Q \leq S \leq P_GJ + C_G
\end{align}
If the settlement S falls below $P_QJ - C_Q$, Quincy will sue. The $P_QJ - C_Q$ determines Quincy's minimum willingness to accept a settlement. But if the settlement amount is too high ($S > P_GJ + C_G$), then Genevieve will refuse to settle. The $P_GJ + C_G$ term determines Genevieve's maximum willingness to pay a settlement. In that case, it is cheaper for Genevieve to go to trial. 

\subsection{Comparative statics}

\begin{figure}
    \centering
    \includegraphics[width=0.8\linewidth]{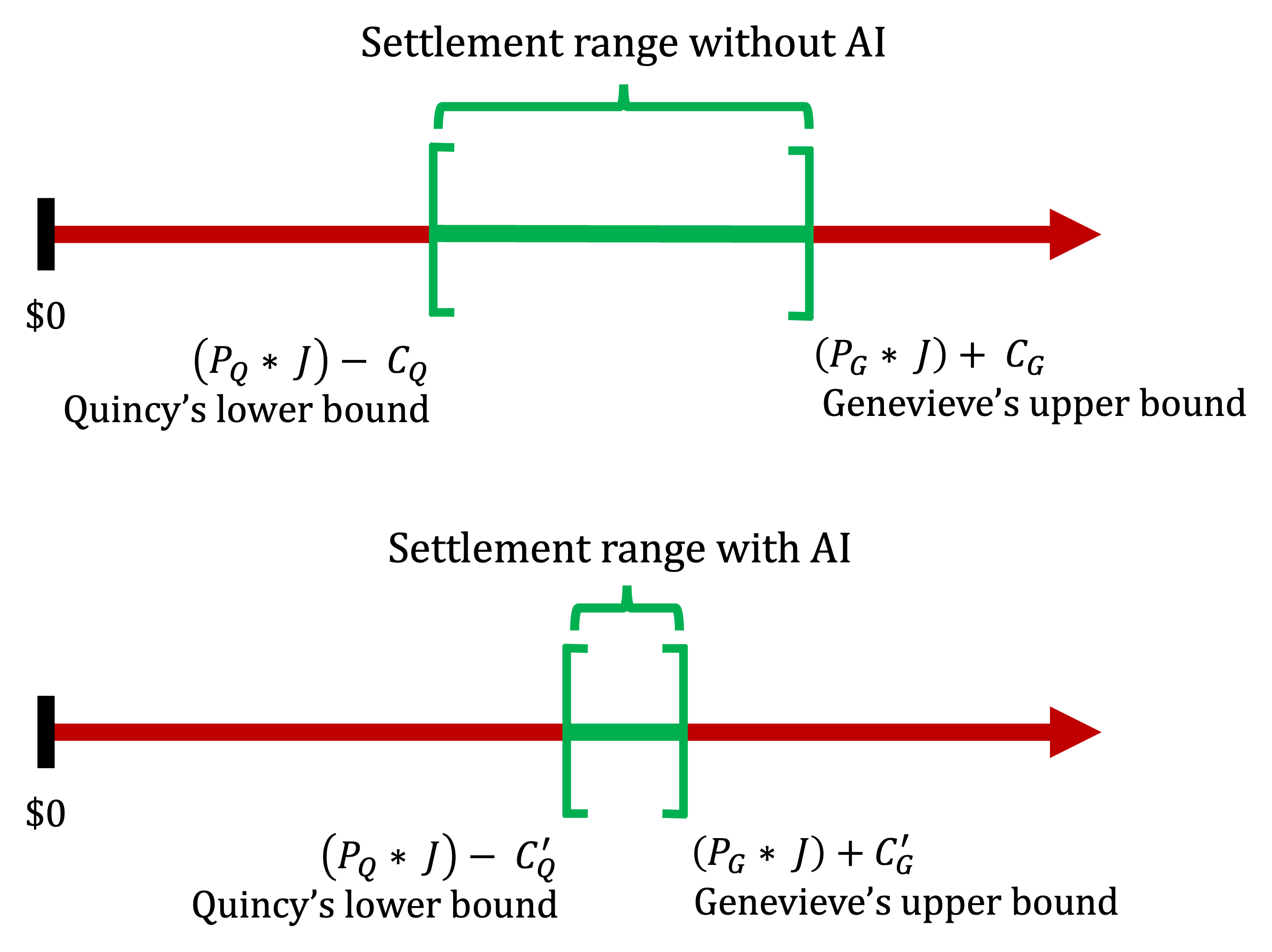}
    \caption{The settlement range}
    \label{fig:setrange}
\end{figure}

A trial is more likely when the cost of going to trial is lower, when Quincy is more optimistic relative to Genevieve, and when the court's judgment ($J$) is larger. Generative AI will reduce the cost of going to trial for both Quincy and Genevieve.

Recall that the litigation effect will reduce the cost of trial attorney services. Thus generative AI will reduce both $C_Q$ and $C_G$, reducing the range within which settlement is possible. The settlement range shrinks because Quincy's minimum willingness to accept gets larger and Genevieve's maximum willingness to pay a settlement gets smaller. By reducing the ``tax" on litigation, generative AI will cause more cases to go to trial rather than be settled.\footnote{Although this paper focuses on the contracting and litigating effect of generative AI, generative AI may be able to predict the outcome of trials and so, in turn, shape how optimistic disputants are. See, for example, \citet{casey2020will} and \citet{gans2024demand}.}

Figure \ref{fig:setrange} shows the settlement range with and without generative AI. In both panels is the continuum of settlement amounts, increasing from left to right. This continuum runs from zero to infinity, reflecting the range of settlement offers that could be offered. The colors denote which settlement amounts are possible.

In the upper panel, if the settlement falls on the red line between zero and $P_QJ - C_Q$, Quincy will go to trial. If the settlement is on the red line to the right of $P_GJ + C_G$, Genevieve will reject the settlement and both parties will go to trial. Settlement is possible only if it falls along the green line that is within the upper and lower bounds. The lower panel of Figure \ref{fig:setrange} shows that generative AI shrinks the settlement range by reducing the cost of litigation.

\subsection{Implications for courts}
While contracts and courts are substitutes, attorneys and courts are complements; they are used together. Ergo when the price of attorneys falls, the demand for its complement, courts, will rise. Following the comparative statics above, the reduction in the cost of hiring attorneys will increase the demand for court services. This is because when the costs of hiring lawyers fall, any disputes that do occur are more likely to be adjudicated in court rather than settled. Ergo we can expect generative AI to cause the rate of litigated disputes to rise.

\section{AI and the evolution of law}

Changes in the degree of contract completeness and increases in the incentive to go to trial will change the long-run evolution of the law. I draw on seminal papers by \citet{rubin1977common} and \cite{priest1977common} that argue that the common law naturally selects for efficient legal rules over time. The reason is simple. As inefficient rules are more likely to be litigated than efficient rules, inefficient rules do not ``survive" in the long run.

\subsection{Why the law evolves}

When disputes or accidents occur, the total costs thereof include both the cost of the harm itself as well as the opportunity cost of resources used to avoid the harm. Ideally, tort law assigns liability to the least-cost avoider. Similarly, contract and property law ought to give the right to use or sell to the highest-valued user. Assigning liability to the least-cost avoider, for example, ensures that precaution is produced at the lowest possible cost. Such an allocation of rights is ``efficient."

``Inefficient" rules imply a less-than-ideal allocation of rights. Inefficient rules allocate accident liability to the highest-cost avoider or the right to act to the lowest-valued user. In accident law, for example, such an allocation is inefficient because any given level of harm prevention is now produced at a higher cost than would otherwise be the case.

\begin{figure}
    \centering
    \includegraphics[width=0.9\linewidth]{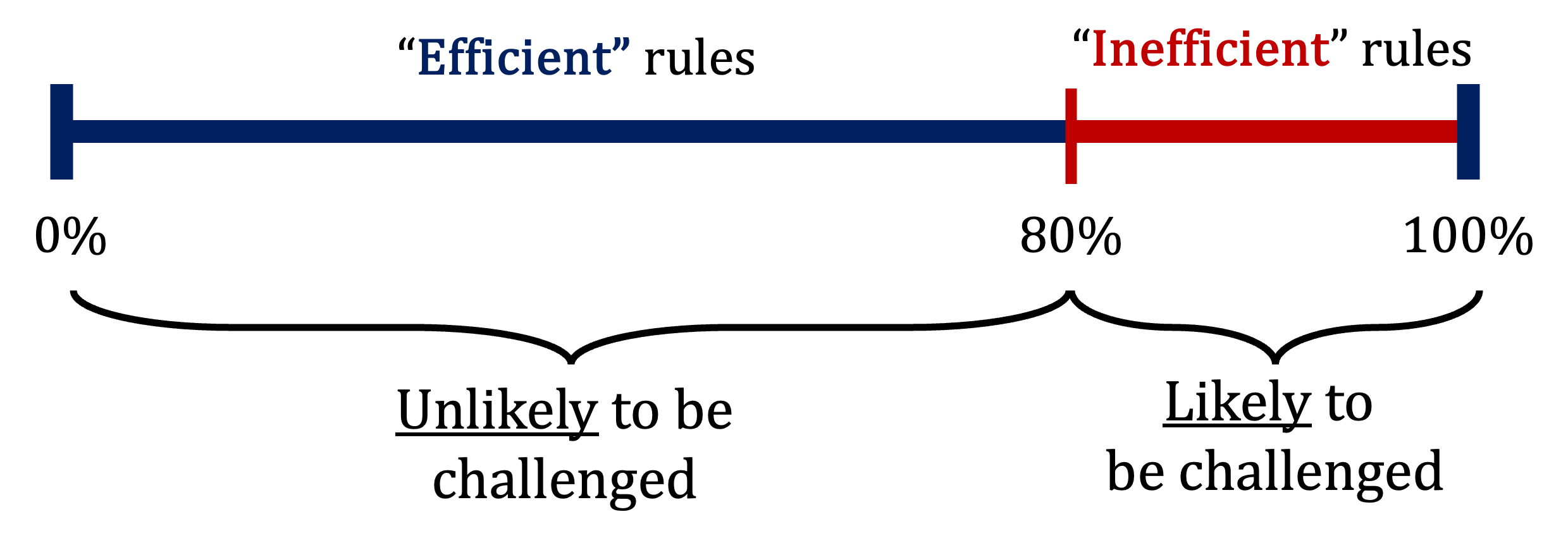}
    \caption{Fraction of the law's rules that are efficient}
    \label{fig:efficientrules}
\end{figure}

\cite{priest1977common} argues that since inefficient rules impose greater costs on the parties subject to them relative to efficient rules, the inefficient rules are more likely to be challenged in court and ultimately overturned as indicated by Figure \ref{fig:efficientrules}. As an example, suppose there are two parties, Quincy and Genevieve. Quincy has a higher cost of avoiding an accident compared to Genevieve. 

Unfortunately for Quincy, the current rule is ``inefficient": Quincy is liable for any accidents that occur. Given his higher costs of avoidance, Quincy spends a great deal on avoiding an accident.\footnote{One could also argue that his higher costs of avoidance will cause Quincy to spend very little on avoidance. In that case, accident costs will be higher as they will occur more often and be more severe. The reasoning below does not change whether Quincy spends too much or too little on avoidance. Quincy will have an incentive to challenge the rule either way. He challenges to either economize on the precaution costs or accident costs.} This need not always be the case, however. If an accident occurs, Quincy can go to court and challenge the law's current allocation of liability. If Quincy loses, he remains liable for all accidents. If Quincy wins, the allocation of liability switches to Genevieve who is then liable for any other accidents that occur. 

Quincy values the right to be free of liability more than Genevieve does. This is true because the resources Quincy spends avoiding an accident determine his maximum willingness to challenge the current rule. The higher are his costs of taking care, the more he is willing to pay to challenge the current liability rule. Quincy expects to save a great deal once liability switches from him to Genevieve. As a result, Quincy has a strong incentive to challenge the rule in court. 

If Quincy succeeds in overturning the rule in court, Genevieve must instead take care in avoiding accidents, which is costly. Genevieve's incentive to challenge the rule is lower than Quincy's because her costs of avoidance are lower. Genevieve's benefits from challenging are lower than Quincy's and so she is less willing to challenge a rule under which she is liable.

The fact that Quincy has a stronger incentive to litigate the inefficient rule than Genevieve's incentive to litigate the efficient rule implies that, in the long run, the right to escape liability will find its way to its higher-valued user: Quincy.\footnote{\cite{priest1977common} stresses that the result holds even if judges decide cases randomly or are biased against efficient assignments of liability.} The conclusion is general. Inefficient rules are not likely to last because they are the ones most likely to be (re-)litigated. According to Priest, 

\begin{quote}
Efficient rules `survive' in an evolutionary sense because they are less likely to be relitigated and thus less likely to be changed. Inefficient rules `perish' because they are more likely to be reviewed and review implies the chance of change whatever the method of judicial decision \citeyearpar[72]{priest1977common}. 
\end{quote}
\noindent Like markets, litigation allows property rights to find their way to their highest-valued users, net of the costs of litigation.

\subsection{Where AI will shape the evolution of law}
Generative AI will impact the rate at which the law evolves via changes in the demand for courts. Contracts may become more or less complete, depending upon the relative elasticity of the marginal cost and benefit curves associated with contractual gap filling. In markets where contracts become more complete, the demand for courts will fall. In markets where contracts become less complete, the demand for courts will rise. Generative AI will also reduce the cost of hiring via the attorney effect, implying an increased demand for courts for disputes that do occur.

The change in demand for courts due to generative AI will not be the same in every area of law. While the rates of evolution in property and contract law are not likely to change, generative AI is likely to accelerate the evolution of tort law. As a result, generative AI will have an uneven effect on the evolution of law.

In property and contract law, parties often have the foresight to write detailed agreements. In these areas of law, the amount of litigation that will occur is unlikely to change much due to the offsetting effects of generative AI on the degree of contract completeness. Although contracts may reduce the cost of creating more complete contracts, they also reduce the benefits thereof by reducing the cost of litigation. 

Generative AI is likely to accelerate the evolution of tort law towards efficiency, however. In tort law, disputes occur without pre-arranged contracts. As a result, the contracting effect, whether it increases or decreases contract completeness on net, is not present. Only the attorney effect is present, implying a reduced cost of litigation and an increased demand for court services. This means that inefficient rules that were previously too costly to litigate can now be challenged, accelerating the replacement of inefficient rules.

\section{Robustness checks}
The results above are robust to both changes in the amount of frivolous litigation and changes in who pays trial costs.

\subsection{Frivolous litigation}
Frivolous litigation is unlikely to upset this result. To illustrate, I draw on Rosenberg and Shavell's \citeyearpar{rosenberg1985model} model of frivolous litigation. The simplified version predicts that the evolution of law will remain unaffected by changes in the incentive to file frivolous lawsuits because frivolous lawsuits never go to trial. 

Suppose that Genevieve may be liable for an accident. There are two victims: Quincy and Oliver. Quincy's case has merit and so the court will rule in his favor against Genevieve. Oliver's case, by contrast, has no merit and the court will rule against him in favor of Genevieve. Although Quincy and Oliver know the merits of their own cases, Genevieve does not.

Consider Oliver's incentive to file a frivolous claim. Hoping to earn a settlement $S$, he moves first and can file suit against Genevieve at cost $f_O$. 

At this early stage, Genevieve can choose how to respond. She can settle, defend herself, or just go to trial without defending herself. If Genevieve settles, she pays S. Oliver then earns $S-f_O$. If Genevieve does not hire an attorney, she is guaranteed to lose in court and must pay a default judgment of $J$. Oliver then gets the benefit of the judgment, less the cost of filing: $J - f_O$. If Genevieve does defend herself, she hires an attorney at cost $d$. At this point, Oliver then can choose to drop the suit or go to trial. As the suit is frivolous, Oliver knows he will lose the trial. His payoff in this case is negative: $- f_O - C_Q$. If he drops the suit, he only pays the filing cost $f_O$. Oliver will always drop the case when Genevieve defends herself.

Quincy has the meritorious case. He has the same options as Oliver. If Genevieve settles, Quincy gets $S-f_Q.$ If Genevieve hires the trial attorney, Quincy can go to trial in which case he will get $J - f_Q - C_Q$, or he can drop in which case he will get $-f_Q$. As Quincy's payoff from going to trial is always greater than the payoff from dropping the suit ($f_Q - C_Q > -f_Q$), Quincy will always go to trial when Genevieve defends herself.

What will Genevieve choose? In Oliver's case, a settlement is always feasible when $f_O \leq S \leq d$. As long as the settlement amount $S$ is less than the cost of hiring the attorney $d$, Genevieve will always settle. Oliver knows this. Oliver will choose to file a frivolous suit as long as $f_O \leq d$. But if $S \leq d$, Genevieve will always go to trial. Oliver knows this and so will never file because the payoff would be negative for him.

Generative AI will reduce both Oliver's costs of filing, $f_O$, and Genevieve's costs of defending herself, $d$. As a result, generative AI may lead to more or fewer frivolous lawsuits. If $f_O$ falls faster than $d$, the settlement range will grow, leading to a larger equilibrium quantity of frivolous filings. If $d$ falls faster than $f_O$, the settlement range will fall, leading to a smaller quantity of frivolous filings.  

Although the number of frivolous filings may change, the rate at which the law evolves will not. This is because Oliver's frivolous suits never go to trial. He always either settles or drops the suit.

\subsection{The English rule}
The acceleration in the evolution of law also happens under the English rule. Under the American rule, detailed in Section \ref{sec:settlementdecision}, Quincy is responsible for his litigation costs and Genevieve pays hers. By contrast, under the English rule, the trial's loser pays everyone's trial costs \citep{cooter1989economic}. The winner pays nothing. For example, if Genevieve loses, she has to pay both her own litigation costs and Quincy's. 

Generative AI has the same effect on the incentive to litigate under the English Rule as it does on the American rule. Even when the loser pays extra, generative AI incentivizes more litigation, increasing the demand for and the use of court services.

Consider the exploding van example above. Once more, Quincy can settle or go to trial. As before, if Quincy offers to settle with Genevieve and Genevieve agrees, Quincy gets a settlement of $S$. Genevieve pays $S$. 

The key difference is in Quincy's relative benefits from winning or losing at trial. If Quincy goes to trial and wins, Quincy gets the judgment of size $J$ but does not have to pay the cost of going to trial. Thus his payoff from winning at trial is simply $J$. Now Genevieve pays the judgment $J$, her costs of going to trial $C_G$, and Quincy's costs of going to trial, $C_Q$. Thus her total cost of losing is $J+C_G+C_Q$.

If, however, Quincy goes to trial and loses, he has to pay both his cost of going to trial, $C_Q$, and Genevieve's cost of going to trial, $C_G$. Thus his total cost of going to trial in this case is $C_Q + C_G$. Genevieve, by contrast, pays nothing so her payoff from winning at trial is zero.

As before, Quincy thinks that his chance of winning at trial is $P_Q$. Quincy's expected benefits of going to trial is therefore
\begin{align}
    P_QJ-(1-P_Q)(C_Q + C_G)
\end{align}

Conversely, Genevieve thinks that the chance that she loses liability at trial is $P_G$, and the probability that she escapes liability is $1-P_G$. Thus her expected costs of going to trial are

\begin{align*}
    P_G(J+ C_G+C_Q)+(1-P_G)(0)
\end{align*}
or
\begin{align}
    P_G(J+ C_G+C_Q).
\end{align}

Quincy will settle when the benefit of settlement is greater than or equal to the expected benefit of going to trial:
\begin{align*}
    S \geq P_Q J-(1-P_Q)(C_Q + C_{G}).
\end{align*}
Quincy will, by contrast, go to trial when the expected benefit of going to trial exceeds the benefit of settling:
    \begin{align*}
    P_QJ-(1-P_Q)(C_Q + C_{G}) > S.
\end{align*}

Genevieve will settle when the cost of settlement is lower than the expected cost of going to trial:
    \begin{align*}
    P_G(J+ C_G+C_Q) \geq S.
\end{align*}
Genevieve will prefer to go to trial when the expected cost of the trial is lower than the cost of the settlement:
    \begin{align*}
    S > P_G(J+ C_G+C_Q).
\end{align*}

Ergo, settlement will occur as long as S is higher than Quincy's expected benefit of trial and is lower than Genevieve's expected cost of going to trial:
\begin{align}
    P_QJ-(1-P_Q)(C_Q + C_{G}) \leq S \leq P_{G}(J+C_Q + C_{G})
    \label{eqn:english}
\end{align}

The logic is the same as in Section \ref{sec:settlementdecision}. If the settlement $S$ falls below $P_QJ-(1-P_Q)(C_Q + C_{G})$, Quincy will sue. And if the settlement amount is too high ($P_G(J+ C_G+C_Q$), then Genevieve will refuse to settle. 

\subsubsection{Comparative statics}

To make comparative statics more clear, we can rearrange Equation \ref{eqn:english}. Settlement is feasible as long as
\begin{align}
    (P_Q-P_{G})(J + C_Q + C_{G}) \leq C_Q + C_{G}
\end{align}
Relative to the American rule, settlement is in general less likely because the left-hand side (LHS) is larger. 

As under the American rule, AI via the ``litigation effect" will cause more trials to occur. Assume that $C_Q$ and $C_{G}$ are continuous, twice differentiable, and are falling in the quality of generative AI. The other parameters are not.

The second partial derivative of the LHS with respect to generative AI is $(P_Q-P_{G})C''_Q$, whereas the second partial derivative of the right-hand side is $C''_Q$. Since $(P_Q-P_{G})C''_Q<C''_Q$, an increase in the costs of litigation will cause the right-hand side to increase more quickly relative to the left-hand side, implying that settlement is more likely as the costs of litigation rise. Therefore the converse states that as generative AI reduces the cost of going to court, settlement will become less likely because the right-hand side will fall more quickly than the right-hand side. As under the American rule, the settlement range will shrink because Quincy's minimum willingness to accept will grow and Genevieve's minimum willingness to pay will fall as the costs of litigation fall. And, as under the American rule, the demand for and use of courts will rise.

Although generative AI has the same kind of effect on the incentive to litigate under the English rule as it does the English rule, the size of the effect is not necessarily the same across the two. The size of the effect depends upon which party is more pessimistic. 

Recall that, under the English rule, the loser pays all trial costs. Generative AI will not change the behavior of a party that is confident that it will win because that party expects that it will not pay the litigation costs. However, generative AI will change the behavior of a pessimistic party. The reason is that they expect to lose and pay all the trial costs. So when the trial costs of the pessimistic party fall, their incentive to go to trial will rise. 

When the defendant is pessimistic, the American rule will witness a larger increase in the incentive to go to trial relative to the English rule. Recall that the defendant has to pay the judgment plus all the trial costs. Trial costs are only a portion of the pessimistic defendant's total costs. So when generative AI reduces the cost of litigation, the pessimistic defendant's total costs fall little in relative terms under the English rule.

But when the plaintiff is pessimistic, the English rule will see more cases go to trial relative to the American rule. In this case, trial costs make up the entire portion of a pessimistic plaintiff's total costs; he does not need to pay a judgment. This means that when generative AI reduces the cost of litigation, the pessimistic plaintiff's total costs fall substantially in relative terms under the English rule. 

To see how, suppose that the width of the settlement range under the American rule, $W_A$ is the upper bound minus the lower bound:
\begin{align*}
    W_A=[P_G J + C_G] - [P_Q J - C_Q]
\end{align*}
or
\begin{align}
    W_A=(P_G - P_Q) J + (C_Q + C_G).
    \label{eqn:widthamericanbefore}
\end{align}
The width of the settlement range under the English rule $W_E$ is 
\begin{align*}
  W_E = [P_G J + P_G (C_Q + C_G)] - [P_Q J - (1 - P_Q)(C_Q + C_G)]
\end{align*}
or
\begin{align}
    W_E = (P_G - P_Q) J + [ (P_G + 1 - P_Q) (C_Q + C_G) ].
    \label{eqn:widthenglishbefore}
\end{align}
Equations \ref{eqn:widthamericanbefore} and \ref{eqn:widthenglishbefore} say that a smaller width ($W_A$ or $W_E$) implies a lower chance of settlement. 

If generative AI reduces everyone's cost of litigation under both rules by $\Delta C$, then Equation \ref{eqn:widthamericanbefore} becomes 
\begin{align}
    W'_A = (P_G - P_Q) J + (C_Q + C_G - 2 \Delta C)
\end{align}
and Equation \ref{eqn:widthenglishbefore} becomes 
\begin{align}
    W'_E = (P_G - P_Q) J + [ (P_G + 1 - P_Q) (C_Q + C_G - 2 \Delta C) ].
    \label{eqn:widthenglishafter}
\end{align}
While the width of the settlement range under the American rule shrinks by $2 \Delta C$, the settlement range under the English rule shrinks by $(P_G + 1 - P_Q)2 \Delta C$. To put the effect of generative AI in relative terms, consider the ratio of the two shrinkages: $\frac{1}{P_G + 1 - P_Q}$. 

If  $\frac{1}{P_G + 1 - P_Q}=1$, then generative AI will shrink the width of the settlement range in both sets of rules by the same amount. This will occur when Genevieve and Quincy are equally optimistic, $P_G = P_Q$. Both sets of rules will have the same-sized increase in the incentive to go to trial. 

If  $\frac{1}{P_G + 1 - P_Q} <1$, then the width of the settlement range will shrink more under the English rule because the decline in Quincy's total costs of going to trial is large. People under the English rule will have a stronger incentive to go to trial relative to people under the American rule when both get access to generative AI. This will occur when Quincy, the plaintiff, is more pessimistic than Genevieve, the defendant ($P_G > P_Q$). 

But if  $\frac{1}{P_G + 1 - P_Q} >1$, then the width of the settlement range will shrink more under the American rule because the decline in Genevieve's total costs of going to trial is small. People under the American rule will have a stronger incentive to go to trial relative to people under the English rule. This will occur when Genevieve, the defendant, is more pessimistic than Quincy ($P_G < P_Q$).

\section{Conclusion}

\begin{figure}[ht]
    \centering
    \includegraphics[width=1\linewidth]{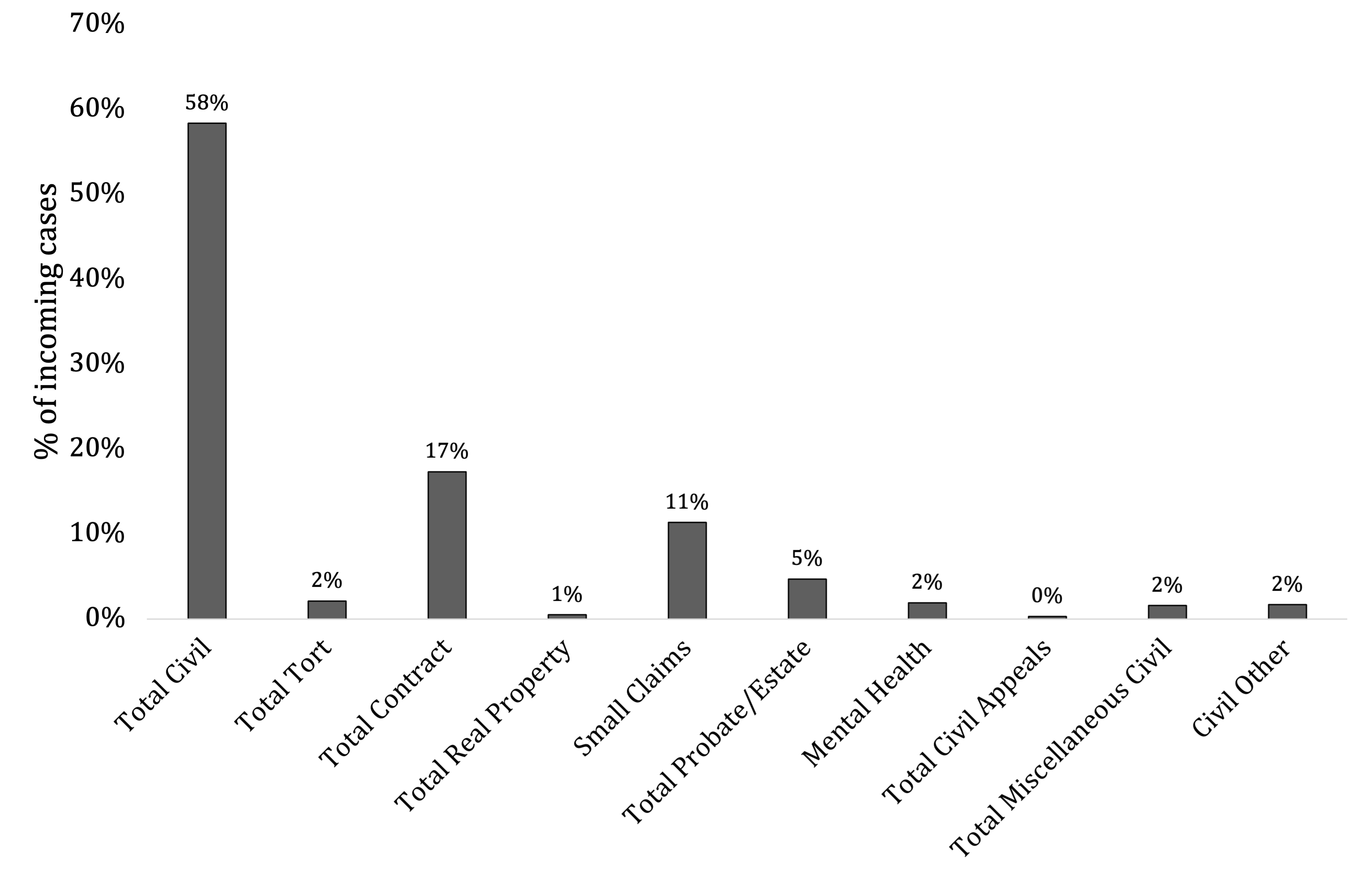}
    \caption{Percentage of incoming cases for state courts, by category}
    \label{fig:incomingcases}
\vspace{-2mm}
    \justify{
            {\footnotesize \textit{Notes: Data is from \cite{gibson2024csp}. Virginia, Mississippi, Kansas, Alabama, and Louisiana were not included in the figure as $\approx 90\%$ of the National Center for State Court's data is missing for those states.}
            }  
	}
\end{figure}

My analysis offers three conclusions pertinent to law, economics, and AI. First, the rate of legal change depends upon both legal and non-legal factors. Until now, past scholarship has stressed non-legal factors such as the Industrial Revolution. I argue that legal factors such as the productivity of attorneys can be just as important. Generative AI is poised to show just that. 

Second, the rate of legal change is sensitive to what attorneys are good at doing. Attorneys may be best known for helping to resolve disputes in court. However, attorneys play an equally important role in avoiding disputes by drafting better contracts. As a result, the total effect of generative AI on the law will depend upon how much better attorneys become in each role.

Third, technological innovation shapes the arc of legal change. Just as increased manufacturing complexity spurred the shift from negligence to strict product liability \citep{landes1989economic}, so too could generative AI spur changes in precedent. Rules of precedent that were, until generative AI, too costly to overturn will be challenged.

For example, the National Center for State Courts \citeyearpar{gibson2024csp} has current data on the incoming caseloads of 46 of the U.S. states. In 2019, tort cases (2\%) were some of the least common incoming cases relative to civil cases (51\%) and contract cases (17\%). Thanks to generative AI, litigation in all areas of tort law will rise. But if areas of tort law with high costs of litigation (such as environmental torts) also have the least efficient rules, then we can expect the most changes in legal precedent to occur there.\footnote{There may be one offsetting effect. If the costs of litigation fall by a flat amount across all areas, then the areas with the highest costs of litigation will decrease less in relative terms. Areas with low litigation costs will decrease more in relative terms. In that case, areas with low litigation costs will see the greatest increase in litigation. This is an example of the so-called ``Alchian-Allen" effect \citep{hummels2004shipping}.}

\newpage
\bibliographystyle{apalike}
\bibliography{References}

\begin{thebibliography}{}

\bibitem[Alarie et~al., 2018]{alarie2018artificial}
Alarie, B., Niblett, A., and Yoon, A.~H. (2018).
\newblock How artificial intelligence will affect the practice of law.
\newblock {\em University of Toronto Law Journal}, 68:106--124.

\bibitem[Albrecht et~al., 2022]{albrecht2022evolution}
Albrecht, B.~C., Hendrickson, J.~R., and Salter, A.~W. (2022).
\newblock Evolution, uncertainty, and the asymptotic efficiency of policy.
\newblock {\em Public Choice}, 192(1):169--188.

\bibitem[Allen, 1998]{allen1998compatible}
Allen, D.~W. (1998).
\newblock Compatible incentives and the purchase of military commissions.
\newblock {\em The Journal of Legal Studies}, 27(1):45--66.

\bibitem[Allen, 2011]{allen2011institutional}
Allen, D.~W. (2011).
\newblock {\em The institutional revolution: Measurement and the economic emergence of the modern world}.
\newblock University of Chicago Press, Chicago.

\bibitem[Allen and Barzel, 2011]{allen2011evolution}
Allen, D.~W. and Barzel, Y. (2011).
\newblock The evolution of criminal law and police during the pre-modern era.
\newblock {\em The Journal of Law, Economics, \& Organization}, 27(3):540--567.

\bibitem[Allen and Leeson, 2015]{allen2015institutionally}
Allen, D.~W. and Leeson, P.~T. (2015).
\newblock Institutionally constrained technology adoption: Resolving the longbow puzzle.
\newblock {\em The Journal of Law and Economics}, 58(3):683--715.

\bibitem[Arbel and Hoffman, 2024]{arbel2024generative}
Arbel, Y. and Hoffman, D.~A. (2024).
\newblock Generative interpretation.
\newblock {\em New York University Law Review}, 99:451.

\bibitem[Arbel et~al., 2024]{arbel2024systemic}
Arbel, Y., Tokson, M., and Lin, A. (2024).
\newblock Systemic regulation of artificial intelligence.
\newblock {\em Arizona State Law Journal}, 56:545.

\bibitem[Arbel, 2024]{arbel2024judicial}
Arbel, Y.~A. (2024).
\newblock Judicial economy in the age of \text{AI}.
\newblock Mimeo.

\bibitem[Batchelder and Freudenberger, 1983]{batchelder1983rational}
Batchelder, R.~W. and Freudenberger, H. (1983).
\newblock On the rational origins of the modern centralized state.
\newblock {\em Explorations in Economic History}, 20(1):1.

\bibitem[Bick et~al., 2024]{bick2024rapid}
Bick, A., Blandin, A., and Deming, D.~J. (2024).
\newblock The rapid adoption of generative \text{ai}.
\newblock Working Paper 32966, National Bureau of Economic Research.

\bibitem[Brynjolfsson et~al., 2023]{brynjolfsson2023generative}
Brynjolfsson, E., Li, D., and Raymond, L.~R. (2023).
\newblock Generative \text{AI} at work.
\newblock Working Paper 31161, National Bureau of Economic Research.

\bibitem[Casey and Niblett, 2020]{casey2020will}
Casey, A.~J. and Niblett, A. (2020).
\newblock Will robot judges change litigation and settlement outcomes?
\newblock Mimeo.

\bibitem[Chiang et~al., 2024]{chiang2024chatbot}
Chiang, W.-L., Zheng, L., Sheng, Y., Angelopoulos, A.~N., Li, T., Li, D., Zhang, H., Zhu, B., Jordan, M., Gonzalez, J.~E., and Stoica, I. (2024).
\newblock Chatbot arena: An open platform for evaluating \text{LLMs} by human preference.
\newblock Mimeo.

\bibitem[Choi et~al., 2023]{choi2023lawyering}
Choi, J.~H., Monahan, A., and Schwarcz, D. (2023).
\newblock Lawyering in the age of artificial intelligence.
\newblock Mimeo.

\bibitem[Cooter and Rubinfeld, 1989]{cooter1989economic}
Cooter, R.~D. and Rubinfeld, D.~L. (1989).
\newblock Economic analysis of legal disputes and their resolution.
\newblock {\em Journal of Economic Literature}, 27(3):1067--1097.

\bibitem[Co{\c{s}}gel et~al., 2012]{cocsgel2012politicaleconomy}
Co{\c{s}}gel, M.~M., Miceli, T.~J., and Rubin, J. (2012).
\newblock The political economy of mass printing: Legitimacy and technological change in the ottoman empire.
\newblock {\em Journal of Comparative Economics}, 40(3):357--371.

\bibitem[Dell'Acqua et~al., 2023]{dell2023navigating}
Dell'Acqua, F., McFowland~III, E., Mollick, E.~R., Lifshitz-Assaf, H., Kellogg, K., Rajendran, S., Krayer, L., Candelon, F., and Lakhani, K.~R. (2023).
\newblock Navigating the jagged technological frontier: Field experimental evidence of the effects of \text{AI} on knowledge worker productivity and quality.
\newblock {\em Harvard Business School Technology \& Operations Management Unit Working Paper}.

\bibitem[Farnsworth, 2004]{farnsworth1990}
Farnsworth, A.~E. (2004).
\newblock {\em Farnsworth on contracts}, volume~2.
\newblock Aspen Law \& Business, New York, 3rd edition.

\bibitem[Fleck and Hanssen, 2024]{fleck2024courts}
Fleck, R.~K. and Hanssen, F.~A. (2024).
\newblock Courts, legislatures, and evolving property rules: Lessons from eminent domain.
\newblock {\em Explorations in Economic History}, 93:101581.

\bibitem[Fon and Parisi, 2003]{fon2003litigation}
Fon, V. and Parisi, F. (2003).
\newblock Litigation and the evolution of legal remedies: A dynamic model.
\newblock {\em Public Choice}, 116(3):419--433.

\bibitem[Fon and Parisi, 2006]{fon2006judicial}
Fon, V. and Parisi, F. (2006).
\newblock Judicial precedents in civil law systems: A dynamic analysis.
\newblock {\em International Review of Law and Economics}, 26(4):519--535.

\bibitem[Friedman, 1984]{friedman1984rights}
Friedman, L.~M. (1984).
\newblock Rights of passage: Divorce law in historical perspective.
\newblock {\em Oregon Law Review}, 63:649.

\bibitem[Gans, 2024]{gans2024demand}
Gans, J.~S. (2024).
\newblock Demand for artificial intelligence in settlement negotiations.
\newblock Working Paper 32685, National Bureau of Economic Research.

\bibitem[Gifford, 2018]{gifford2018technological}
Gifford, D.~G. (2018).
\newblock Technological triggers to tort revolutions: steam locomotives, autonomous vehicles, and accident compensation.
\newblock {\em Journal of Tort Law}, 11(1):71--143.

\bibitem[Gilson, 1984]{gilson1984value}
Gilson, R.~J. (1984).
\newblock Value creation by business lawyers: Legal skills and asset pricing.
\newblock {\em The Yale Law Journal}, 94(2):239--313.

\bibitem[Gould, 1973]{gould1973economics}
Gould, J.~P. (1973).
\newblock The economics of legal conflicts.
\newblock {\em The Journal of Legal Studies}, 2(2):279--300.

\bibitem[Hummels and Skiba, 2004]{hummels2004shipping}
Hummels, D. and Skiba, A. (2004).
\newblock Shipping the good apples out? an empirical confirmation of the \text{A}lchian-\text{A}llen conjecture.
\newblock {\em Journal of Political Economy}, 112(6):1384--1402.

\bibitem[Landes, 1971]{landes1971economic}
Landes, W.~M. (1971).
\newblock An economic analysis of the courts.
\newblock {\em The Journal of Law and Economics}, 14(1):61--107.

\bibitem[Landes and Posner, 1989]{landes1989economic}
Landes, W.~M. and Posner, R.~A. (1989).
\newblock An economic analysis of copyright law.
\newblock {\em The Journal of Legal Studies}, 18(2):325--363.

\bibitem[Leeson and Pierson, 2017]{leeson2017economic}
Leeson, P.~T. and Pierson, J. (2017).
\newblock Economic origins of the no-fault divorce revolution.
\newblock {\em European Journal of Law and Economics}, 43:419--439.

\bibitem[McGinnis and Pearce, 2013]{mcginnis2013great}
McGinnis, J.~O. and Pearce, R.~G. (2013).
\newblock The great disruption: How machine intelligence will transform the role of lawyers in the delivery of legal services.
\newblock {\em Fordham Law Review}, 82:3041.

\bibitem[Noy and Zhang, 2023]{noy2023experimental}
Noy, S. and Zhang, W. (2023).
\newblock Experimental evidence on the productivity effects of generative artificial intelligence.
\newblock {\em Science}, 381(6654):187--192.

\bibitem[Parisi, 2001]{parisi2001genesis}
Parisi, F. (2001).
\newblock The genesis of liability in ancient law.
\newblock {\em American Law and Economics Review}, 3(1):82--124.

\bibitem[Parisi, 2002]{parisi2002entropy}
Parisi, F. (2002).
\newblock Entropy in property.
\newblock {\em American Journal of Comparative Law}, 50:595.

\bibitem[Peng et~al., 2023]{peng2023impactaideveloperproductivity}
Peng, S., Kalliamvakou, E., Cihon, P., and Demirer, M. (2023).
\newblock The impact of \text{AI} on developer productivity: Evidence from github copilot.
\newblock Mimeo.

\bibitem[Posner, 1972]{posner1972theory}
Posner, R.~A. (1972).
\newblock A theory of negligence.
\newblock {\em The Journal of Legal Studies}, 1(1):29--96.

\bibitem[Priest, 1977]{priest1977common}
Priest, G.~L. (1977).
\newblock The common law process and the selection of efficient rules.
\newblock {\em The Journal of Legal Studies}, 6(1):65--82.

\bibitem[Rosenberg and Shavell, 1985]{rosenberg1985model}
Rosenberg, D. and Shavell, S. (1985).
\newblock A model in which suits are brought for their nuisance value.
\newblock {\em International Review of Law and Economics}, 5(1):3--13.

\bibitem[Rubin, 2014]{rubin2014printing}
Rubin, J. (2014).
\newblock Printing and protestants: an empirical test of the role of printing in the reformation.
\newblock {\em Review of Economics and Statistics}, 96(2):270--286.

\bibitem[Rubin, 1977]{rubin1977common}
Rubin, P.~H. (1977).
\newblock Why is the common law efficient?
\newblock {\em The Journal of Legal Studies}, 6(1):51--63.

\bibitem[Shavell, 1982]{shavell1982suit}
Shavell, S. (1982).
\newblock Suit, settlement, and trial: A theoretical analysis under alternative methods for the allocation of legal costs.
\newblock {\em The Journal of Legal Studies}, 11(1):55--81.

\bibitem[\text{National Center for State Courts}, 2024]{gibson2024csp}
\text{National Center for State Courts} (2024).
\newblock Court statistics project \text{STAT}.
\newblock \url{www.courtstatistics.org}.

\bibitem[Toner-Rodgers, 2024]{toner2024artificial}
Toner-Rodgers, A. (2024).
\newblock Artificial intelligence, scientific discovery, and product innovation.
\newblock {\em Mimeo}.

\bibitem[Tu et~al., 2023]{tu2023artificial}
Tu, S.~S., Cyphert, A., and Perl, S.~J. (2023).
\newblock Artificial intelligence: Legal reasoning, legal research and legal writing.
\newblock {\em Minnesota Journal of Law Science \& Technology}, 25:105.

\bibitem[Villasenor, 2023]{villasenor2023generative}
Villasenor, J. (2023).
\newblock Generative artificial intelligence and the practice of law: Impact, opportunities, and risks.
\newblock {\em Minnesota Journal of Law Science \& Technology}, 25:25.

\bibitem[Zheng et~al., 2023]{zheng2023lmsys}
Zheng, L., Chiang, W.-L., Sheng, Y., Li, T., Zhuang, S., Wu, Z., Zhuang, Y., Li, Z., Lin, Z., Xing, E.~P., et~al. (2023).
\newblock Lmsys-chat-1m: A large-scale real-world llm conversation dataset.
\newblock Mimeo.

\end{thebibliography}

\end{document}